\documentclass[sigconf, nonacm]{acmart}

\usepackage{algorithm}
\usepackage{algpseudocode}
\usepackage{subcaption}

\begin{document}
\title{Rateless Bloom Filters: Set Reconciliation for Divergent Replicas with Variable-Sized Elements}

\author{Pedro Gomes}
\affiliation{%
  \institution{University of Porto}
  \city{Porto}
  \state{Portugal}
}
\email{pedromgomes29@gmail.com}

\author{Carlos Baquero}
\affiliation{%
  \institution{University of Porto}
  \city{Porto}
  \state{Portugal}
}
\email{cbm@fe.up.pt}

\begin{abstract}
Set reconciliation protocols typically make two critical assumptions: they are designed for fixed-sized elements and they are optimized for when the difference cardinality, $\mathbf{d}$, is very small. When adapting to variable-sized elements, the current practice is to synchronize fixed-size element digests. However, when the number of differences is considerable, such as after a network partition, this approach can be inefficient. Our solution is a two-stage hybrid protocol that introduces a preliminary Bloom filter step, specifically designed for this regime. The novelty of this approach, however, is in solving a core technical challenge: determining the optimal Bloom filter size without knowing $\mathbf{d}$. Our solution is the Rateless Bloom Filter (RBF), a dynamic filter that naturally adapts to arbitrary symmetric differences, closely matching the communication complexity of an optimally configured static filter without requiring any prior parametrization. Our evaluation in sets of variable-sized elements shows that for Jaccard indices below $85\%$, our RBF-IBLT hybrid protocol reduces the total communication cost by up to over $\mathbf{20\%}$ compared to the state-of-the-art.
\end{abstract}

\maketitle

\section{Introduction}
Set reconciliation is a fundamental problem in distributed systems. Two replicas, Alice and Bob, hold sets \(S_A\) and \(S_B\) of elements, respectively. Their goal is to communicate efficiently to find the differences between their sets so that both replicas can obtain the set union, \(S_A \cup S_B\). This problem has wide-ranging applications, including cloud storage for synchronizing files between replicas, distributed databases for anti-entropy protocols, and blockchain networks for transaction dissemination.

While the problem is often simplified in the literature to focus on sets of fixed-sized elements, real-world synchronization is more complex, frequently involving variable-sized elements such as files, Bitcoin transactions, and database records. This makes handling variable-sized elements a more general and practical problem than what is typically considered. The common approach to adapt fixed-size algorithms to this more general problem is to first create fixed-sized digests (e.g., hashes) of the elements and then apply set reconciliation to these digests. The mismatched digests are then used to identify the variable-sized elements that must be exchanged.

In this work, we present an evaluation of state-of-the-art set reconciliation methods applied to fixed-sized digests of variable-sized elements. We also introduce a new two-stage algorithm that combines Bloom filters with rateless Invertible Bloom Lookup Tables (IBLTs). We demonstrate that for a considerable number of differences, the addition of a preliminary Bloom filter step significantly reduces communication overhead when compared to existing methods. Our method is particularly effective for scenarios of moderate similarity, which are critical after major network events like blackouts or extended replica downtime.

A core challenge with this two-stage approach is how to configure the Bloom filter, as the optimal configuration depends on the unknown cardinality of the set difference. We solve this by introducing rateless Bloom filters, inspired by rateless IBLTs. In our method, nodes continuously stream small Bloom filter with a single hash function, and the receiver signals the sender to stop when the optimal point of communication has been reached. This technique leverages prior work on Bloom joins by Mullin \cite{rateless_bloomjoin}(1990) but is adapted for the set reconciliation problem. As we will show, our method achieves a communication cost very close to that of an optimally configured static Bloom filter, without any need for a priori set difference cardinality estimations.

In summary, we make the following contributions:
\begin{itemize}
    \item We propose a rateless Bloom filter approach, an adaptation of Mullin's multi-filter Bloomjoin for the set reconciliation problem.
    \item We evaluate rateless Bloom filters against standard Bloom filters and show that our approach follows the communication cost of an optimally configured static Bloom filter without requiring any a priori knowledge of the set differences.
    \item We provide a novel evaluation of state-of-the-art set reconciliation algorithms for variable-sized elements, demonstrating that for Jaccard indices below $85\%$, our approach transmits up to over $20\%$ less data than the state-of-the-art algorithm with the lowest communication cost.
\end{itemize}

\section{Motivation}
While set reconciliation is a well-studied problem, state-of-the-art algorithms typically operate under two restrictive assumptions: first, that the symmetric difference cardinality, $d$, is very small compared to the total set sizes; and second, that all set elements are of a fixed size. The common practice of adapting these algorithms for variable-sized elements by hashing them first fundamentally alters their properties. 

While for state-of-the-art methods in fixed-sized element set reconciliation the elements are sent embedded in the transmitted data structures, for variable-sized elements this is not the case. The reconciliation protocol is applied to fixed-size digests, which only serve to identify the differing elements. The actual variable-sized elements must then be transmitted in a separate step. As such, for variable-sized elements and unlike for fixed-size elements, both state-of-the-art set reconciliation algorithms and Bloom filters serve the same purpose: to identify the elements to be exchanged. The question then becomes which is more space-efficient for this identification phase. For considerable symmetric differences, a Bloom filter can be more efficient. For example, the communication cost of using 64-bit digests with PinSketch is at least $d \times 64$ bits, while a Bloom filter's communication cost is usually just a few bits per element (e.g., $n \times 5$). Depending on how small $d$ is compared to $n$, Bloom filters can be more space-efficient than PinSketch in identifying the differing elements.

While very high similarity (small $d$) is expected during routine anti-entropy operations, we argue that scenarios with moderate similarity are not only relevant but critical. These are the precise conditions under which our method provides the most benefit: when the differences are smaller than the whole state, but large enough that it makes sense to use Bloom filters. This sweet spot exists after major events like network partitions, such as the 2025 Iberian Peninsula blackout which lasted for over ten hours, or extended replica downtime, whether due to planned datacenter maintenance or the inherent churn of nodes in P2P networks. In these situations, an efficient set reconciliation protocol is crucial for quickly bringing disconnected replicas up to date.

A core limitation of using Bloom filters for reconciliation is the existence of false positives, which prevent a guaranteed reconciliation. Prior work, such as Graphene\cite{graphene} and ConflictSync\cite{conflictsync}, has demonstrated that combining Bloom filters with more robust reconciliation protocols can result in lower communication costs. However, the optimal configuration of the Bloom filter is dependent on the unknown set difference cardinality. While Graphene handles this by leveraging its specific use case (block dissemination) to make assumptions about set differences, more general applications lack this a priori knowledge. ConflictSync shows that the optimal Bloom filter size increases with the number of differences but does not provide a mechanism for configuring the filter. This exposes a critical research gap: a rateless method in which the Bloom filter size naturally adapts to the set difference cardinality without relying on prior knowledge or estimations.

Additionally, Graphene and ConflictSync were evaluated within their specialized domains (blockchain transaction dissemination and CRDT synchronization, respectively), rather than as a general set reconciliation protocol. As such, they are compared against the state-of-the-art methods within their given domains, and are not evaluated as a generic set reconciliation solution against state-of-the-art set reconciliation methods.

Set reconciliation methods are also typically evaluated for sets of fixed-size elements, stating that applying these algorithms to variable-sized elements is just a question of synchronizing the fixed-size digests and using those to obtain the corresponding elements. However, no prior work has provided a comprehensive evaluation of state-of-the-art set reconciliation solutions on sets of variable-sized elements.

Similar to the hybrid protocols like Graphene and ConflictSync, we introduce a set reconciliation protocol that leverages the combined strengths of Bloom filters and IBLTs. However, we address the critical research gap by introducing Rateless Bloom Filters (RBFs). RBFs naturally adapt to arbitrary symmetric differences with performance closely matching an optimally configured Bloom filter, entirely removing the need for prior estimation or parametrization of $\mathbf{d}$. Furthermore, we benchmark this novel protocol against state-of-the-art set reconciliation protocols in a comprehensive evaluation focused on the crucial, yet neglected, scenario of variable-sized elements.

\section{Background}

\subsection{Bloom filters}

A Bloom filter \cite{bloom} is a space-efficient randomized data structure for approximate set membership queries. It represents a set of $n$ elements using a bit array of size $m$, initially with all bits set to 0. The structure employs $k$ independent hash functions, each mapping an element to a position within the bit array.

\begin{figure}
\centering
\includegraphics[width=0.9\columnwidth]{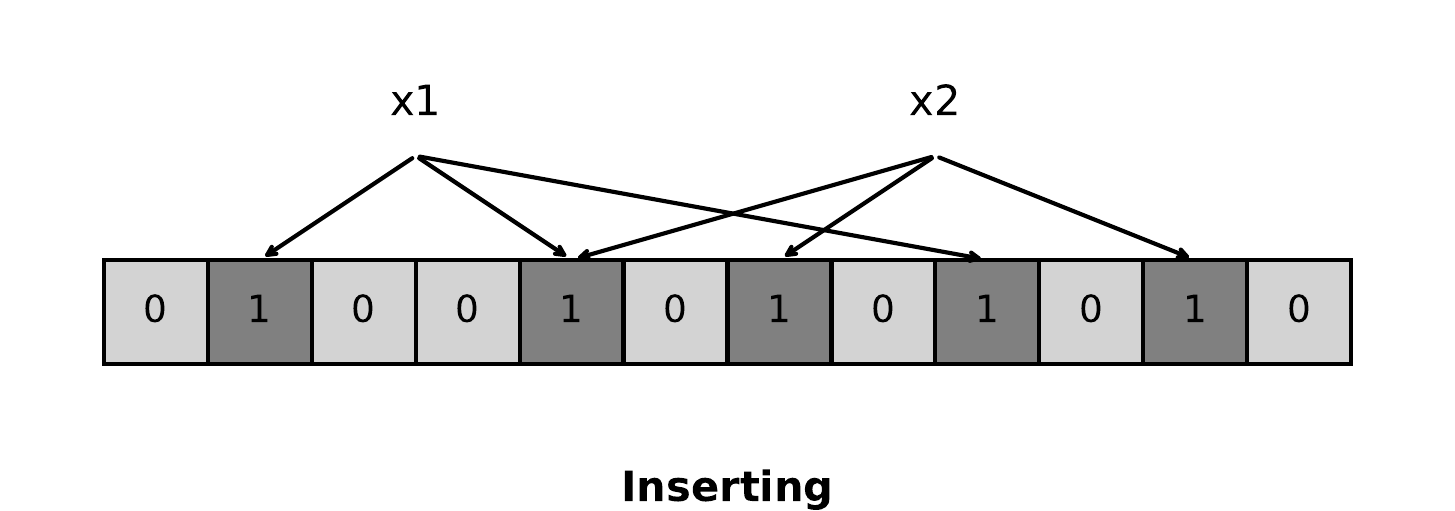}
\caption{Insertion into a standard Bloom filter ($m=12, k=3$). Elements $x_1$ and $x_2$ are inserted.}
\label{fig:bf_insert}
\end{figure}

To insert an element, the bits at the $k$ positions determined by the hash functions are set to 1. As illustrated in Figure~\ref{fig:bf_insert}, with $m=12$ and $k=3$, the insertion of elements $x_1$ and $x_2$ sets specific bits to 1. An element's membership is queried by checking if all $k$ corresponding bits are set to 1. If any of these bits is 0, the element is provably not in the set, ensuring no false negatives. However, multiple insertions can set all the bits for an element that was never added, leading to a false positive.

\begin{figure}
\centering
\includegraphics[width=0.9\columnwidth]{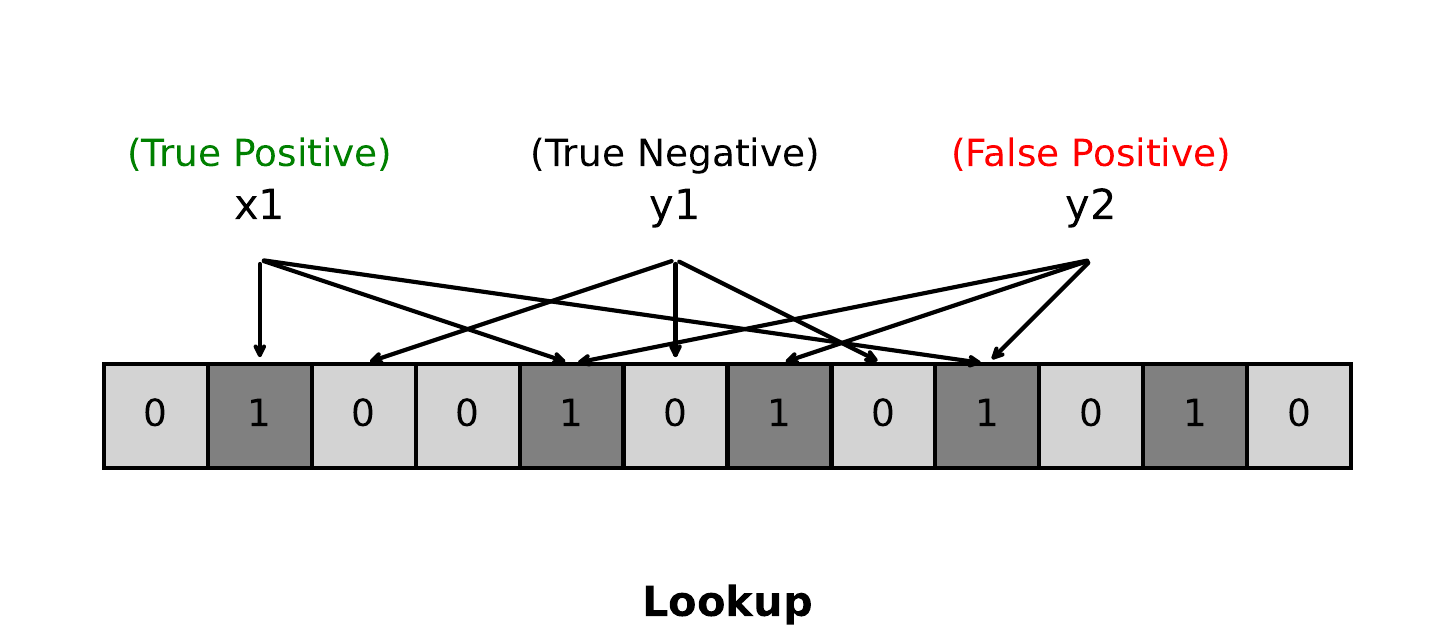}
\caption{Lookup in a standard Bloom filter. $x_1$ is a true positive, $y_1$ is a true negative, and $y_2$ is a false positive.}
\label{fig:bf_lookup}
\end{figure}

Figure~\ref{fig:bf_lookup} demonstrates the lookup process. The element $x_1$, which was previously inserted, correctly yields a positive result. The element $y_1$, never inserted, correctly yields a negative result because at least one of its corresponding bits is 0. However, element $y_2$, also never inserted, results in a false positive because the bits it maps to were all set to 1 by the insertions of $x_1$ and $x_2$.

A variant, the Partitioned Bloom filter\cite{partitioned_bf}, partitions the bit array into $k$ equal-sized blocks of $m/k$ bits. Each of the $k$ hash functions then maps an element exclusively to a position within one of these specific partitions. 
This structure can be viewed as an AND composition of $k$ independent single-hash filters. This composite nature, where the large filter is derived from smaller filters, is fundamental to our approach.


\subsection{Optimal Bloom filter Semijoins}
In a distributed database system, a join between two relations often needs to be performed across different nodes. A naive approach for this is for one node to send its entire relation to the other, where the join is then computed locally. However, it is inefficient to transmit tuples for which there is no join partner on the other side. A semijoin is a technique that reduces a relation to contain only tuples whose join key values find a match on the other relation, thereby ensuring that no useless tuples are transmitted. This problem can be framed as a set membership test: a site needs to send a tuple only if its join key belongs to the set of join keys of the other relation. A naive approach is to send the entire set of distinct join keys from one of the relations, but this is space-inefficient. A better, more space-efficient idea is to use Bloom filters. This approach is called a Bloomjoin~\cite{static_bloomjoin}. False positives can be handled by simply transmitting all the tuples that test positive against the filter, which will necessarily include all the true positives. The cost of false positives is simply an additional communication cost.

The central problem is how to configure the Bloom filter's false positive rate (FPR). If the join keys have high similarity, the Bloom filter is counterproductive as most tuples will test positive anyway. If the join keys have a low similarity, the Bloom filter is a very good idea and should be configured with a low FPR. For intermediate cases, the optimal FPR varies depending on the set difference, a value that is hard to know a priori.

The idea in Mullin's 1990 work~\cite{rateless_bloomjoin} was to avoid this problem by not sending one large, static Bloom filter with a potentially inadequate FPR. Instead, the sender transmits a sequence of small Bloom filters, and the receiver, after processing each one, determines the optimal point to terminate the transmission and signals the sender to stop.

The receiver's role is to decide whether the optimal communication point has been reached. With each small Bloom filter, some number of false positives are revealed as true negatives. This number of eliminated tuples decreases with each additional filter. The receiver, knowing the average size of a tuple and the size of each small Bloom filter, can compute the number of tuples that could be transmitted using the same number of bits. Eventually, the number of tuples that become true negatives by receiving the latest filter will be smaller than the number of tuples that could have been transmitted with the same number of bits. At this point, it would have been more efficient to simply transmit the revealed false positives than to send the new Bloom filter. The receiver notifies the sender to stop, having reached a near-optimal communication cost without any a priori knowledge of the set difference.

\subsection{Rateless Set Reconciliation}
\subsubsection{Invertible Bloom Lookup Tables (IBLTs)}
An Invertible Bloom Lookup Table (IBLT)~\cite{iblt} is a probabilistic data structure that extends the functionality of a traditional Bloom filter. While a Bloom filter can only test for the presence of an element, an IBLT is designed to allow for the reconstruction of the original set of elements with high probability.


Each cell in an IBLT is a tuple of three fields:
\begin{itemize}
    \item \textbf{idSum}: The bitwise XOR sum of the IDs of all elements that map to this cell.
    \item \textbf{hashSum}: The bitwise XOR sum of the hashes of all elements that map to this cell.
    \item \textbf{count}: The number of elements that map to this cell.
\end{itemize}
Elements are inserted into an IBLT by applying a set of $k$ independent hash functions, similar to a Bloom filter. For each of the $k$ resulting cell indices, the element's ID and hash are XORed into the cell's \textit{idSum} and \textit{hashSum}, respectively, and the \textit{count} is incremented. Element removal uses the exact same XOR operations but decrements the \textit{count}. The use of XOR ensures that adding and then removing an element leaves a cell's \textit{idSum} and \textit{hashSum} unchanged.

A cell is considered pure if it contains exactly one element. This can be verified with overwhelming probability by checking if the hash of the \textit{idSum} matches the \textit{hashSum}. This check fails with overwhelming probability for cells containing two or more elements.

When a pure cell is found, the element is identified and peeled away by removing its contribution from all cells it maps to. This peeling process can create new pure cells, leading to a cascading decoding effect. The process continues until all elements are recovered (when all cells become empty) or until no more pure cells can be found, at which point decoding fails.

\begin{figure}[h]
    \centering
    \includegraphics[width=0.85\columnwidth]{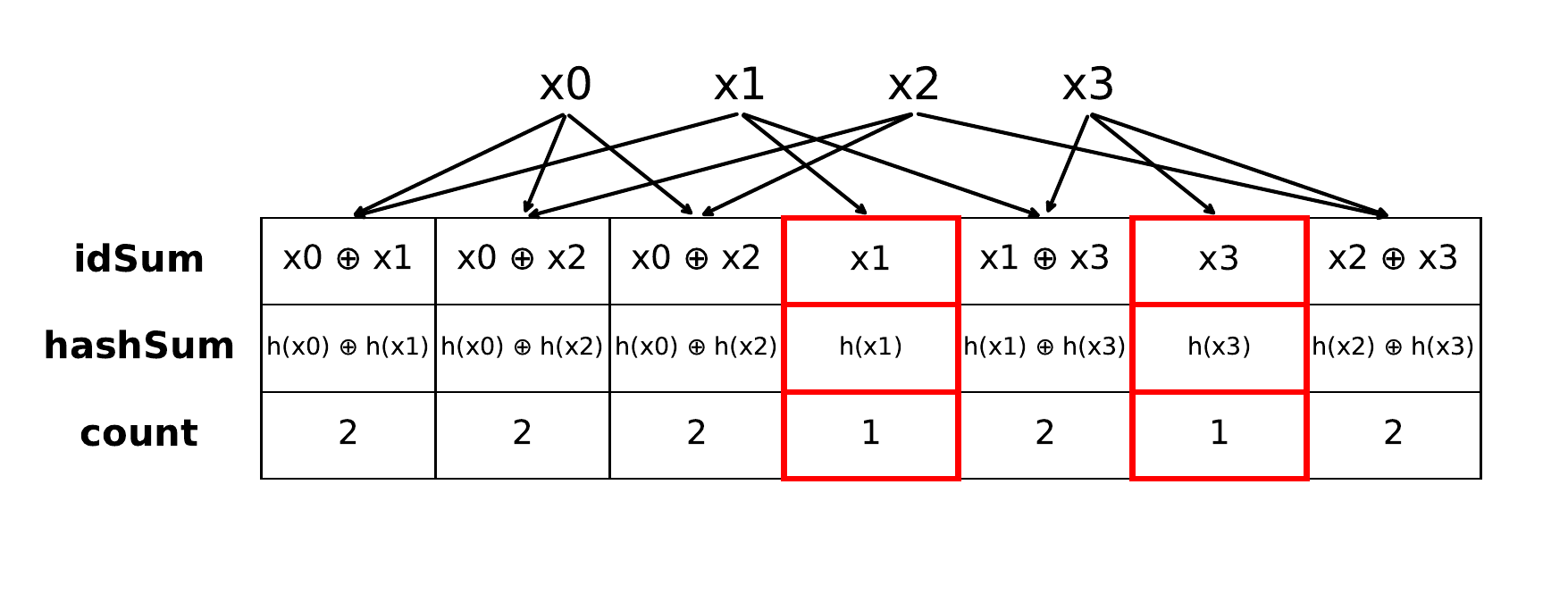}
    \caption{An example IBLT.}
    \label{fig:iblt}
\end{figure}

As illustrated in Figure \ref{fig:iblt}, an IBLT is used to encode a set with elements $x_0, x_1, x_2, x_3$. The cells highlighted in red are pure, as their \textit{count} is 1, and the hash of their \textit{idSum} matches the \textit{hashSum}. This confirms that elements $x_1$ and $x_3$ are in the set. After identifying $x_1$ and $x_3$, the elements can be removed from the cells to which they map, which in turn causes other cells to become pure (e.g., the first and last cell), allowing a cascading decoding process to recover the remaining elements.

\subsubsection{Set reconciliation with IBLTs}
IBLTs provide an efficient mechanism for set reconciliation ~\cite{iblt_reconciliation}. The sender constructs an IBLT $\text{IBLT}_A$ from its set and transmits it to the receiver. The receiver similarly computes $\text{IBLT}_B$ from its local set. Both are then combined to construct a new IBLT representing the symmetric difference, $\text{IBLT}_{B \Delta A}$, by performing a cell-wise combination:
\begin{align*}
    \text{idSum}_{B \Delta A}[i] &= \text{idSum}_B[i] \oplus \text{idSum}_A[i] \\
    \text{hashSum}_{B \Delta A}[i] &= \text{hashSum}_B[i] \oplus \text{hashSum}_A[i] \\
    \text{count}_{B \Delta A}[i] &= \text{count}_B[i] - \text{count}_A[i]
\end{align*}
Due to the properties of XOR, any element present in both sets will cancel out, leaving the combined IBLT with only the XOR sums and counts of elements unique to either set.

The receiver then decodes this difference IBLT. A key insight is that the `count' field in the combined IBLT indicates the origin of an element in a pure cell: a count of 1 signifies an element exclusive to the receiver's set ($B \setminus A$), while a count of -1 indicates an element exclusive to the sender's set ($A \setminus B$). The peeling algorithm proceeds as before, with a modification: when an element is recovered with a count of -1, its contribution is added to the IBLT to remove its effect, whereas for a count of 1, its contribution is subtracted.

\begin{figure}
    \centering
    \includegraphics[width=0.85\columnwidth]{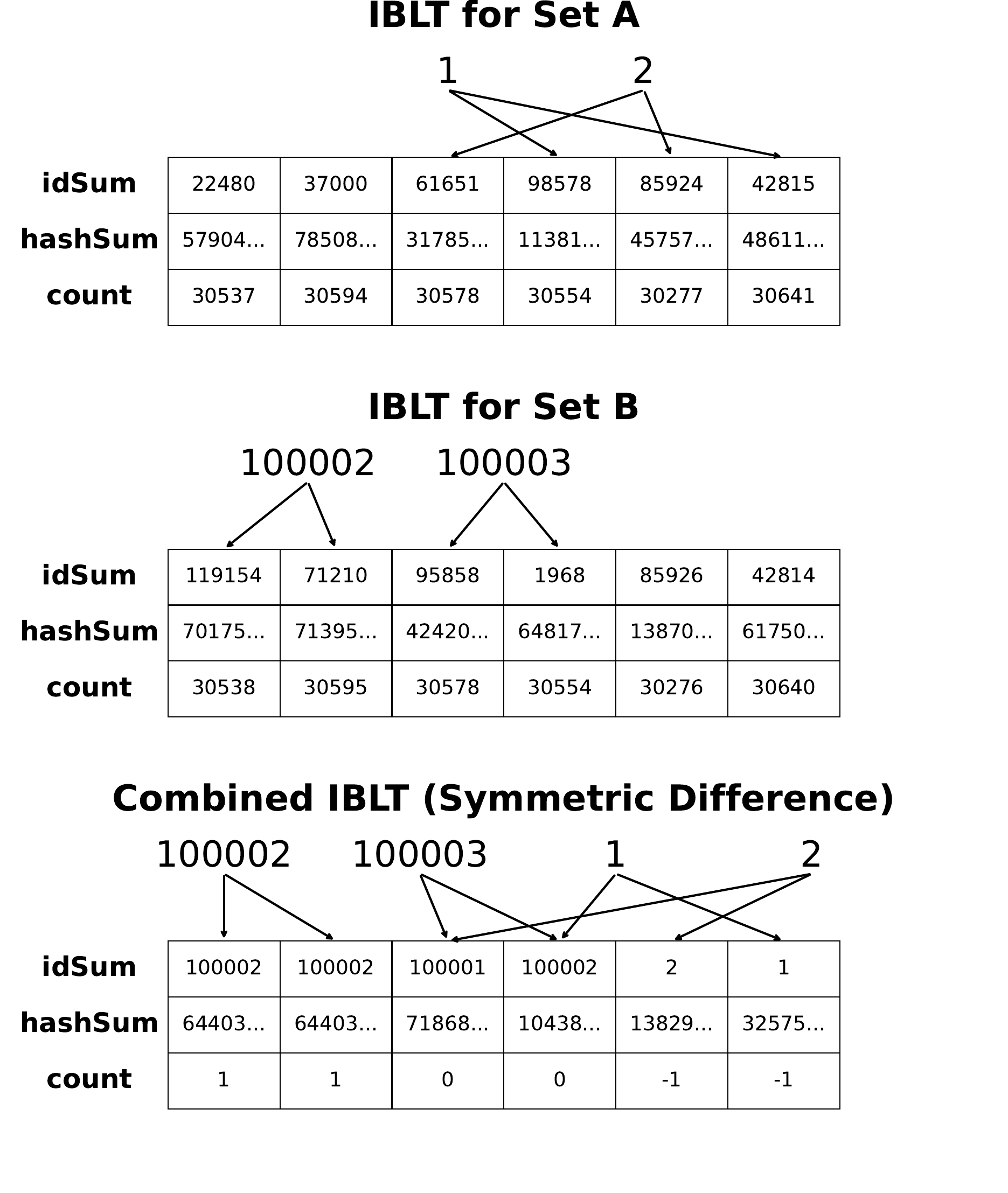}
    \caption{Set reconciliation with IBLTs.}
    \label{fig:iblt_reconciliation}
\end{figure}

As depicted in Figure \ref{fig:iblt_reconciliation}, two IBLTs, $\text{IBLT}_A$ and $\text{IBLT}_B$, are used to reconcile two large sets. Set $A$ contains elements from 1 to 100001, and Set $B$ contains elements from 3 to 100003. This example highlights the space-efficiency of IBLTs. Despite the large size of the original sets (approximately 100,000 elements each), the IBLTs are sized in proportion to the cardinality of the symmetric difference, which in this case is small ($d = 4$), consisting of elements 1, 2, 100002, and 100003. The IBLTs shown in the figure have only 6 cells.

The combined IBLT at the bottom of the figure represents the symmetric difference. The peeling algorithm is then applied. Cells with index 0, 1, 4, and 5 are pure, allowing for the identification and recovery of elements 1, 2, and 100002. From their respective `count` values, we can determine that 1 and 2 are exclusive to the sender (count of -1), while 100002 is exclusive to the receiver (count of 1). After peeling away the identified elements, only 100003 remains in the combined IBLT, which can then be recovered.

\subsubsection{Set reconciliation with rateless IBLTs}
The concept of rateless set reconciliation~\cite{rateless_set_reconciliation} provides an elegant solution to the problem of an unknown symmetric difference size. Instead of a fixed-size IBLT, a party can generate a virtually endless stream of IBLT cells. The opposing party simply consumes this stream until it has received enough cells to successfully decode the symmetric difference.

This is achieved by changing the way we map elements to cells such that we can generate an infinite sequence: instead of using $k$ hash functions, this approach uses a pseudo-random number generator (PRNG). The PRNG, seeded with an element's hash, deterministically generates a sequence of cell mappings for that element. By streaming these cells, the receiver can process them as they arrive. Since both parties use the same PRNG seeded by the same element hashes, the cell mappings are identical on both sides, ensuring that common elements cancel out as expected.

The key insight is that the PRNG can be designed to produce mappings that follow a specific probability distribution. Theoretical analysis and Monte Carlo simulations show that with an appropriate distribution, the number of cells required for decoding is directly proportional to the size of the symmetric difference, $d$. On average, the number of cells needed is approximately $1.72d$ for small $d$ and approaches $1.35d$ as $d$ becomes large.

This rateless approach eliminates the need for any prior estimation of $d$, simplifying the protocol, reducing latency, and guaranteeing that the communication overhead is only as large as necessary to resolve the actual set difference.

\section{Adapting Mullin's Cost Model for Set Reconciliation}\label{sec:rbf_stopping_criterion}
To address the problem of selecting an optimal static Bloom filter size for an unknown symmetric difference $d$, we propose a rateless Bloom filter approach inspired by rateless IBLTs. Our protocol defines a conceptual, infinite partitioned Bloom filter and continuously streams its partitions (also referred to as slices) from the sender to the receiver. The receiver processes this stream and, at each step, evaluates a cost-based decision to terminate the transmission.

In Mullin's pioneering work on Bloom joins, the communication was turn-based. The sender would generate and transmit a small Bloom filter, configured with a single hash function and an optimal density of $1/\ln(2)$ bits per element. The receiver would then process this filter, decide whether to stop, and if not, request another.

This concept of small, one-hash filters can be directly seen as a variant of a partitioned Bloom filter. Each $k$ partitions (or slices) transmitted in sequence form a complete partitioned Bloom filter with $k$ hash functions. This allows our approach to leverage the extensive work done on this Bloom filter variant.

In our work, the sender does not wait for a confirmation before sending additional slices. Instead, it continuously generates and streams them until it receives a signal to stop. This design significantly decreases latency by avoiding the round-trip delay between the sender and receiver. The trade-off is a small, potential communication overhead if slices are transmitted unnecessarily due to network latency between the receiver's decision to stop and the sender receiving that signal.

The core of our method is the adaptation of Mullin's cost-based decision model for the set reconciliation problem. The total communication cost is the sum of the costs to identify the elements exclusive to each replica's set:
$$ \text{Total Cost} = \text{Cost}(A \setminus B) + \text{Cost}(B \setminus A) $$
The cost for a single-direction identification is composed of the Bloom filter size and the cost of transmitting the remaining false positives:
$$ \text{Cost}(A \setminus B) = |\text{BF}_B| +  C_{elem} \times |\text{FP}_{A \setminus B}| $$
and
$$ \text{Cost}(B \setminus A) = |\text{BF}_A| +  C_{elem}\times |\text{FP}_{B \setminus A}| $$
The $C_{elem}$ term represents the communication overhead for a single element in the final reconciliation step. For a protocol like PinSketch, this is the size of a single digest (e.g., 64 bits). For rateless IBLTs, it is the size of an IBLT cell multiplied by the average number of cells needed per element. The receiver continuously computes the cost-benefit of each additional Bloom filter partition. The stream is stopped when the number of new true negatives that are detected by a new partition becomes less than the number of elements that could be reconciled by the set reconciliation protocol using an equivalent number of bits.

\section{Hybrid Rateless Set Reconciliation}
The description of Rateless Bloom filters so far has been generic, as they can be combined with many different set reconciliation protocols. In this work, we combine rateless Bloom filters with rateless IBLTs, resulting in a protocol which we call Hybrid Rateless Set Reconciliation. The reason for this alignment is simple: any other state-of-the-art set reconciliation protocol (like PinSketch) requires an estimate of the symmetric difference cardinality, $d$. This would mean that after exchanging the rateless Bloom filters, we would need to obtain an estimate for the number of false positives (the new $d'$). In contrast, rateless IBLTs require no such estimate and naturally adapt to the true value of $d'$. As the entire purpose of this work is to eliminate the need for set difference cardinality estimations, we believe these two algorithms align nicely to create a fully self-adapting protocol.

As was already explained, the estimated cost per false positive is the size of an IBLT cell multiplied by the expected number of cells per difference. 
We use $1.35$ as the expected number of cells per difference in the cost calculation, as this is the value the ratio converges to. We denote the cost per element in the final reconciliation phase as $C_{elem}$.

The full synchronization protocol is shown in Algorithm~\ref{alg:hybrid-rateless-sync}. The protocol is executed in three distinct phases: two sequential Bloom filter-based identification phases followed by a final, precise IBLT-based reconciliation phase.

\subsection{Phase 1: Unidirectional RBF Streaming}

The protocol initiates with replica $A$ continuously generating and streaming Bloom filter slices. Following the established optimal parameterization for Bloom filters\cite{bloom}, the total filter size $m$ required for maximum information density is $m = (n \cdot k) / \ln(2)$, where $n$ is the set size and $k$ is the number of hash functions. Adopting the settings used by Mullin \cite{rateless_bloomjoin}, we configure each slice with a single hash function ($k=1$) and a slice size $m$ defined by the following optimal value: $m = n / \ln(2)$.

Each slice, indexed by $i$, is then generated using the function \texttt{generateSlice}$(S, i)$, which produces a Bloom filter representation of set $S$ using a single hash function. Specifically, each element $x \in S$ is inserted into the slice by setting the bit at position $H(x \mathbin\Vert i) \bmod m$, where $H$ is a hash function and $\mathbin\Vert$ denotes concatenation. This design avoids the need for both replicas to agree in advance on a fixed set of hash functions, as the slice index effectively acts as a hash function selector. Conceptually, an infinite sequence of Bloom filter slices is defined for a given set, and replica $A$ continuously streams growing prefixes of this sequence.

Replica B uses these slices to iteratively partition its local set $S_B$ into $S_B^{\text{com}}$ (suspected common elements that pass the membership test) and $S_B^{\text{TN}}$ (elements that are confirmed not to be in $S_A$, i.e. true negatives). Initially, all elements are considered suspected common. With each new slice, the `partition` function is invoked to refine this categorization. The function \texttt{partition}($S_{B}^{com}$, $\text{RBF}_{S_A}[i]$) tests all elements in a set against the received Bloom filter slice, returning a tuple of two subsets: the first for elements that pass the membership test (and remain in the `suspected common` set), and the second for elements that fail the test (and are moved to the `true negatives` set).

The second subset corresponds to the newly revealed negatives. The stream from replica $A$ is terminated based on a cost-based stopping criterion. When the condition $|S_B^{\text{new\_TN}}| < m / C_{elem}$ is verified, it signifies that the cost of a new $m$-bit slice ($m$) exceeds the cost of reconciling those elements directly using set reconciliation (the number of newly revealed true negatives, $|S_B^{\text{new\_TN}}|$, multiplied by the cost to reconcile an element in the final phase, $C_{elem}$). At this point, replica $A$ is signaled to stop.

\subsection{Phase 2: Bidirectional Partitioning}
Upon receiving the stop signal, replica $B$ constructs and streams a new RBF over its `suspected common` elements to replica $A$. This enables $A$ to apply the same partitioning procedure to its own set $S_A$, producing subsets $S_A^{\text{com}}$ and $S_A^{\text{TN}}$. This bidirectional exchange ensures that both replicas have identified a common set of elements and a set of local true negatives.

\subsection{Phase 3: Rateless IBLT Reconciliation}
The final phase uses rateless set reconciliation to identify and resolve the false positives within the sets of suspected common elements, $S_A^{\text{com}}$ and $S_B^{\text{com}}$. Reconciliation is performed over the fixed-size hashes of the elements. For details on how to efficiently generate, decode, and verify IBLTs, consult the background section.

When replica $B$ decodes the IBLT difference and identifies the hashes of the elements it is missing from $A$, it cannot recover the original elements locally. Therefore, it transmits these hashes back to $A$, which then responds with the actual elements corresponding to those hashes. These elements are the false positives from the RBF phases. The algorithm concludes with each replica performing a final set union operation, incorporating the true negatives and false positives from its peer, thereby ensuring full set synchronization.

\begin{algorithm}
\caption{
Hybrid Rateless Set Synchronization:
}\label{alg:hybrid-rateless-sync}
\begin{algorithmic}[1]
    \State \textbf{durable state:} $S_A, S_B$
    \State \textbf{local state at B:} $S_B^{\text{com}} \gets S_B$, $S_B^{\text{TN}} \gets \emptyset$
    \State \textbf{local state at A:} $S_A^{\text{com}} \gets S_A$, $S_A^{\text{TN}} \gets \emptyset$

    \While{B has not signaled stop}
        \State $\text{send}_{A,B}(\text{RBFStream}, i, \text{generateSlice}(S_A, i))$
        \State $i \gets i + 1$
    \EndWhile

    \Procedure{OnReceive$_{A,B}$}{$\text{RBFStream}, i, \text{RBF}_{S_A}[i]$}
        \State $(S_B^{\text{com}}, S_B^{\text{new\_TN}}) \gets \text{partition}(S_B^{\text{com}}, \text{RBF}_{S_A}[i])$
        \State $S_B^{\text{TN}} \gets S_B^{\text{TN}} \cup S_B^{\text{new\_TN}}$
        \State $m \gets |S_A|/\ln(2)$
        \If{$|S_B^{\text{new\_TN}}| < m / C_{elem}$}
            \While{A has not signaled stop}
                \State $\text{send}_{B,A}(\text{RBFStream}, j, \text{generateSlice}(S_B^{\text{com}}, j))$
                \State $j \gets j + 1$
            \EndWhile
        \EndIf
    \EndProcedure

    \Procedure{OnReceive$_{B,A}$}{$\text{RBFStream}, j, \text{RBF}_{S_B^{\text{com}}}[j]$}
        \State $(S_A^{\text{com}}, S_A^{\text{new\_TN}}) \gets \text{partition}(S_A^{\text{com}}, \text{RBF}_{S_B^{\text{com}}}[j])$
        \State $S_A^{\text{TN}} \gets S_A^{\text{TN}} \cup S_A^{\text{new\_TN}}$
        \State $m \gets |S_B^{\text{com}}|/\ln(2)$
        \If{$|S_A^{\text{new\_TN}}| < m / C_{elem}$}
            \State $H_A \gets \{ \text{hash}(s) \mid s \in S_A^{\text{com}} \}$
            \While{B has not signaled stop}
                \State $\text{send}_{A,B}(\text{IBLTStream}, k, \text{generateIBLTCell}(H_A, k))$
                \State $k \gets k + 1$
            \EndWhile
        \EndIf
    \EndProcedure

    \Procedure{OnReceive$_{A,B}$}{$\text{IBLTStream}, k, \text{IBLT}_{H_A}[k]$}
        \State $H_B \gets \{ \text{hash}(s) \mid s \in S_B^{\text{com}} \}$
        \State ${\text{IBLT}_{\Delta}}[k] \gets \text{generateIBLTCell}(H_B, k) \oplus \text{IBLT}_{H_A}[k]$
        \If{${\text{IBLT}_{\Delta}}$ is decodable}
            \State $(H_{A\setminus B}, H_{B \setminus A}) \gets \text{decode}({\text{IBLT}_{\Delta}})$
            \State $S_B^{\text{FP}} \gets \{ s \mid s \in S_B^{\text{com}} \land \text{hash}(s) \in H_{B \setminus A} \}$
            \State $\text{send}_{B,A}(\text{FinalUpdate}, S_B^{\text{TN}}, S_B^{\text{FP}}, H_{A\setminus B})$
        \EndIf
    \EndProcedure

    \Procedure{OnReceive$_{B,A}$}{$\text{FinalUpdate}, S_B^{\text{TN}}, S_B^{\text{FP}}, H_{A\setminus B}$}
        \State $S_A^{\text{FP}} \gets \{ s \mid s \in S_A^{\text{com}} \land \text{hash}(s) \in H_{A\setminus B} \}$
        \State $S_A \gets S_A \cup S_B^{\text{TN}} \cup S_B^{\text{FP}}$
        \State $\text{send}_{A,B}(\text{FinalUpdate}, S_A^{\text{TN}}, S_A^{\text{FP}})$
    \EndProcedure
    
    \Procedure{OnReceive$_{A,B}$}{$\text{FinalUpdate}, S_A^{\text{TN}}, S_A^{\text{FP}}$}
        \State $S_B \gets S_B \cup S_A^{\text{TN}} \cup S_A^{\text{FP}}$
    \EndProcedure
\end{algorithmic}
\end{algorithm}

\section{Evaluation}
This section presents a comprehensive experimental evaluation of our proposed rateless Bloom filter protocol. For the sake of reproducibility, the full source code and all data used in our experiments have been made publicly available.

For clarity in data presentation, the y-axis scales of our figures are adjusted dynamically to best illustrate the characteristics of each metric. Therefore, readers should note that the scales may vary across plots when making comparisons.

\subsection{Experimental Setup}
This section details the methodology used to evaluate our protocol's performance against state-of-the-art set reconciliation algorithms. Our entire evaluation was conducted in a controlled simulation environment to ensure reproducibility and accurate measurement of both communication and computational overheads. All experiments were executed on a single machine to eliminate network latency and variability, with communication costs calculated by measuring the serialized byte size of all inter-replica messages.

A fundamental principle of set reconciliation protocols is that they must eventually transmit the differing elements to each replica, which constitutes a theoretical minimum communication cost. All state-of-the-art methods, with the exception of full state transfer, are designed to achieve this. As such, the total amount of mismatched state data transmitted is a constant across all algorithms for a given set of differences. The only variable component is the size of the \textbf{metadata} required to identify the changed elements. Our evaluation, therefore, focuses on minimizing this variable cost.

The distribution of element sizes does not affect the metadata cost. Bloom filters are inherently independent of element size, as they operate by setting a fixed number of bits based on an element's hash rather than storing the element itself. Similarly, other state-of-the-art set reconciliation algorithms operate on fixed-size digests of elements. The size of these digests is chosen to minimize hash collisions with high probability, regardless of the original element size distribution. While different element size distributions exist in real-world scenarios and could affect the overall impact of the metadata cost, the relative comparison of algorithms remains unaffected. For these experiments, we used a uniform distribution over the interval $[5, 80]$ bytes.

Instructions on how to reproduce all experiments can be found in our public repository.\footnote{\texttt{https://github.com/pedrogomes29/rbf}}

\subsubsection{Workload and Methodology}
For each experiment, we generated two replica sets, $S_A$ and $S_B$, each with a fixed cardinality of 100,000 distinct items. The differences between these sets were distributed symmetrically such that $|S_A \setminus S_B| = |S_B \setminus S_A|$. The total number of differing elements, $d = |S_A \setminus S_B| + |S_B \setminus S_A|$, was precisely controlled to achieve a target similarity $s$, defined by the Jaccard index, which is standard in the literature:
$$
s = J(S_A, S_B) = \frac{|S_A \cap S_B|}{| S_A \cup S_B|}
$$

In our experiments, we plotted all metrics against the number of differing elements $d$. The values of $d$ were chosen such that the similarity between the sets ranges from 0\% to 100\%, in 5\% increments. 

To ensure the statistical validity of our results, for each data point (i.e., each number of differing elements), we executed the experiment 30 times with different randomly generated sets. The plots present the average value for each metric, with a shaded surface area showing the $\pm 1$ standard deviation.

\subsubsection{Metrics}
We quantified the performance of each algorithm using the following key metrics:
\begin{itemize}
    \item \textbf{Transmitted Metadata:} The total data size of all information required to detect the differing elements.
    \item \textbf{Total Transmitted Data:} The sum of the transmitted metadata and the serialized state data of the differing elements.
    \item \textbf{Communication Overhead:} The ratio of the total transmitted data to the theoretical minimum communication cost (i.e., the cost of simply transmitting the differing elements). This metric is undefined for a similarity of 100\%, as the minimum cost is zero.
    \item \textbf{Encoding Time:} The time required for a replica to construct the data structures for transmission.
    \item \textbf{Decoding Time:} The time required for a replica to process the received data structures and identify the mismatched elements.
    \item \textbf{Computation Time:} The sum of the encoding and decoding times, representing the total computational cost of the reconciliation protocol.
\end{itemize}

\subsection{Static Bloom Filters vs Rateless Bloom Filters}
\begin{figure*}
    \centering
    \begin{subfigure}{0.45\linewidth}
        \centering
        \includegraphics[width=\linewidth]{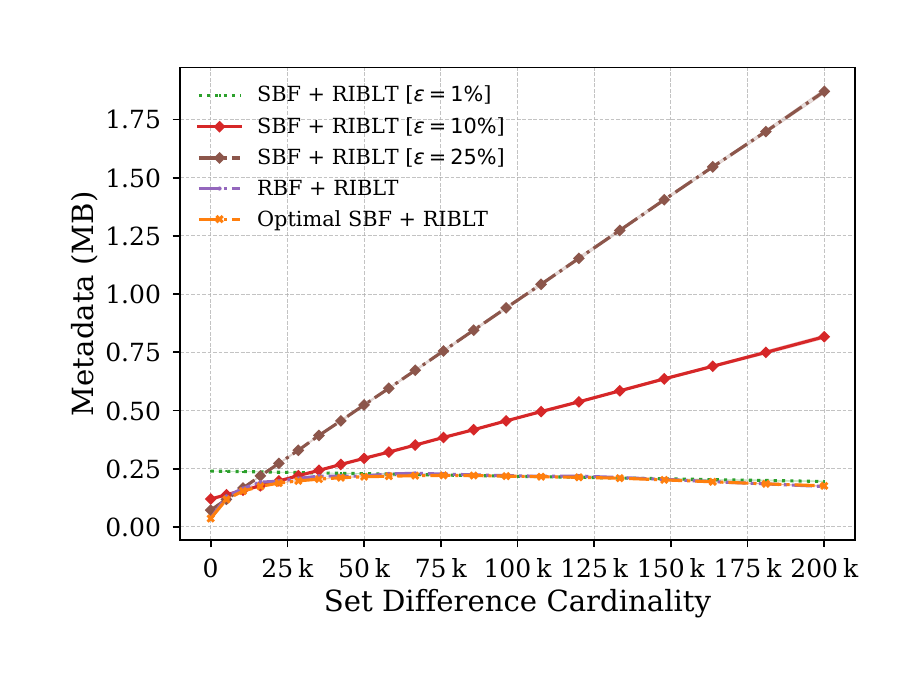}
        \caption{Transmitted metadata}
        \label{fig:sbf_vs_rbf_metadata}
    \end{subfigure}%
    \begin{subfigure}{0.45\linewidth}
        \centering
        \includegraphics[width=\linewidth]{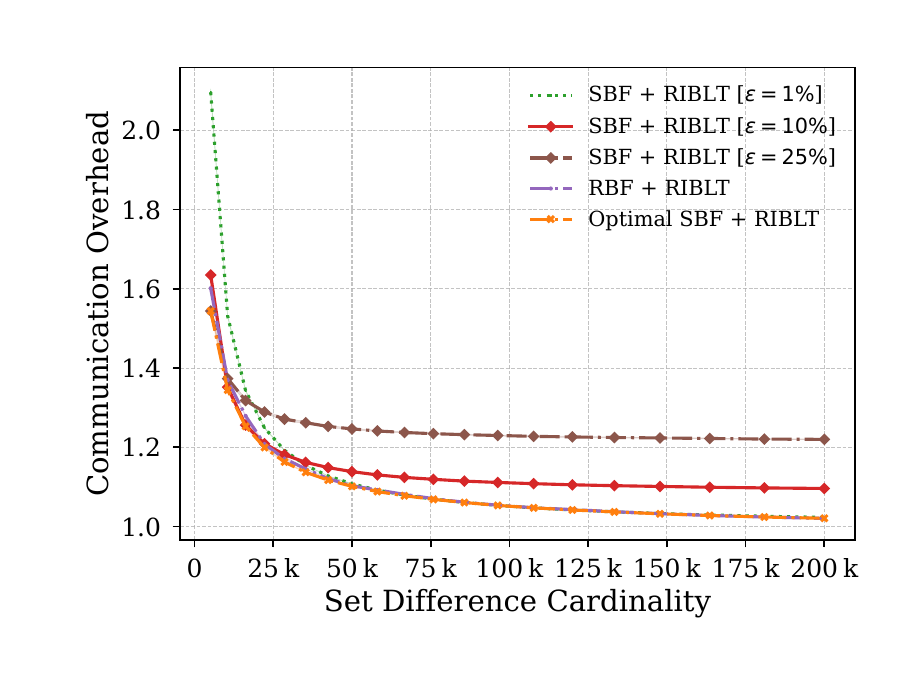}
        \caption{Communication overhead}
        \label{fig:sbf_vs_rbf_communication_overhead}
    \end{subfigure}

    \par 
    
    \begin{subfigure}{0.45\linewidth}
        \centering
        \includegraphics[width=\linewidth]{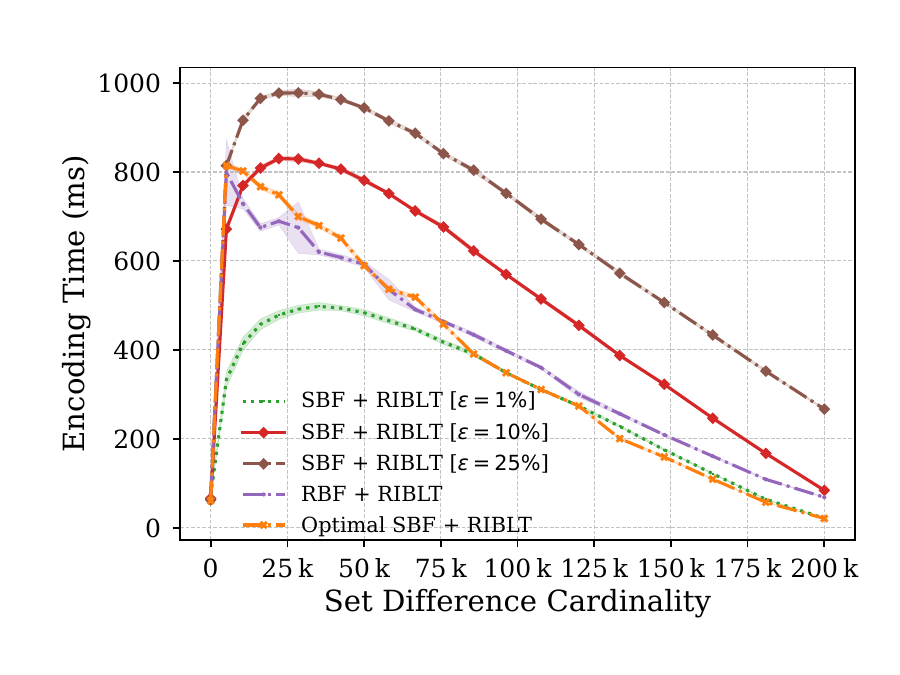}
        \caption{Encoding Time}
        \label{fig:sbf_vs_rbf_encoding_time}
    \end{subfigure}%
    \begin{subfigure}{0.45\linewidth}
        \centering
        \includegraphics[width=\linewidth]{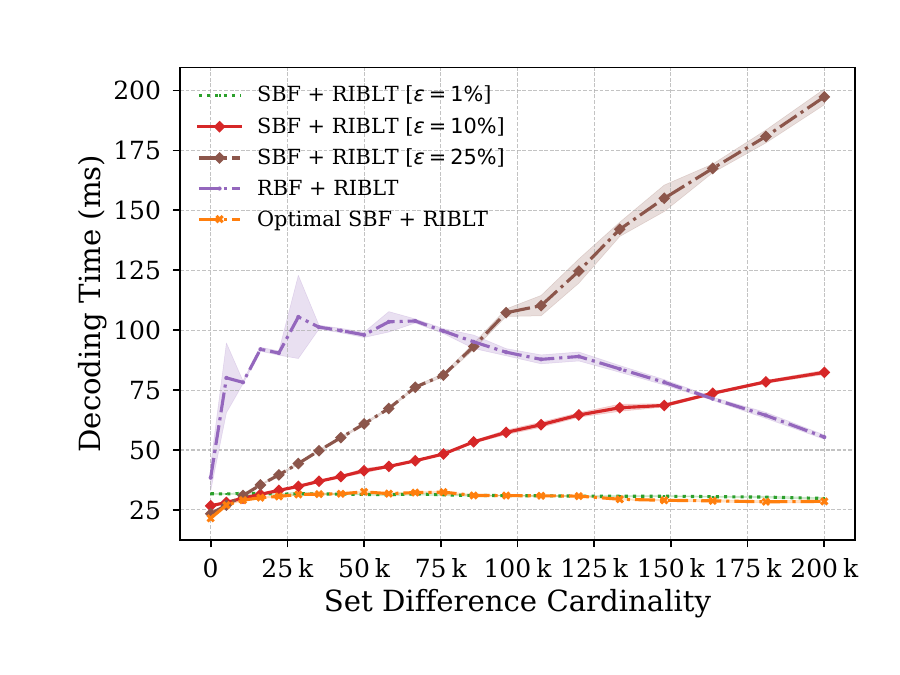}
        \caption{Decoding Time}
        \label{fig:sbf_vs_rbf_decoding_time}
    \end{subfigure}
    \caption{Static Bloom Filters vs. Rateless Bloom filters}
    \label{fig:sbf_vs_rbf}
\end{figure*}

In this experiment, we evaluate the communication and computational costs of our proposed rateless Bloom filter (RBF) against two classes of static Bloom filters (SBFs). The first class consists of SBFs with fixed false positive rates ($\epsilon$), representing common configurations an application might blindly choose. The second, denoted as Optimal SBF, serves as a theoretical performance baseline. This optimal baseline was obtained by generating all SBF configurations with false positive rates from $0.5\%$ to $50\%$ in $0.5\%$ increments and selecting the configuration with the minimum communication cost for each level of set difference cardinality. This comparison allows us to evaluate the practical efficiency of our adaptive RBF against both potentially unoptimized static configurations and the (full knowledge) theoretical lower bound of a static approach.

\subsubsection{Transmitted Metadata and Communication Overhead}
The transmitted metadata graph in Figure \ref{fig:sbf_vs_rbf_metadata} illustrates the core advantage of a rateless design. As depicted, the rateless Bloom filter (RBF) consistently traces the theoretical minimum of an optimal static Bloom filter (SBF). This demonstrates that our RBF closely approximates the performance of a standard Bloom filter whose false positive rate is perfectly tuned for each specific similarity level, without any need for parameterization or prior estimation of set difference cardinalities. In contrast, while a static Bloom filter's metadata cost is fixed regardless of set difference, its rigid design incurs a significant penalty for inadequately chosen false positive rates. For completely dissimilar sets (maximum difference cardinality), a Bloom filter with a high false positive rate of $\epsilon=25\%$ transmits over $7\times$ more metadata compared to our rateless filter. Conversely, for highly similar states (minimal differences), a Bloom filter configured for a low false positive rate of $\epsilon=1\%$ transmits over $4\times$ more metadata.

These differences in transmitted metadata directly impact the communication overhead. As shown in Figure \ref{fig:sbf_vs_rbf_communication_overhead}, our rateless Bloom filter consistently follows the most efficient static configuration for any given point in the set difference space. Static configurations, on the other hand, face significant overheads when not optimized for the specific level of set divergence. This highlights the practical limitations of static designs in real-world scenarios where set differences can vary widely.

\subsubsection{Computation Time}
The encoding and decoding times of the protocols, depicted in Figures \ref{fig:sbf_vs_rbf_encoding_time} and \ref{fig:sbf_vs_rbf_decoding_time}, reveal a more nuanced perspective on performance. While communication overhead is a primary metric, the computational efficiency determines practical feasibility.

\paragraph{Encoding Time}
The encoding time graph in Figure \ref{fig:sbf_vs_rbf_encoding_time} shows a non-linear behavior for the static Bloom filter configurations, characterized by an initial increase, a peak, and a subsequent decrease. This can be explained by the computational complexity of the reconciliation process. Replica A constructs a Bloom filter over its entire set $S_A$. Replica B constructs a Bloom filter over the elements of its set $S_B$ that test positive. Both replicas then build a Rateless IBLT over their respective positive partitions. The dominant cost for this process is the construction of the Rateless IBLT, which is an $O(n \log d)$ operation. Here, $n$ represents the number of elements in the positive partition of the local set, and $d$ is the number of false positives remaining in both sets. This complexity arises from inserting $n$ elements into a Rateless IBLT of size $O(d)$, with each insertion being an $O(\log d)$ operation.

The distinct shape of the encoding time curves is the result of two competing dynamics. As the number of differences increases, the size of the set intersection decreases, which causes the number of true positives to decrease, lowering $n$. However, some of these true positives which are now negative will test positive, resulting in an increase in the number of false positives. Initially, the logarithmic growth in $d$ dominates the linear decrease in $n$, causing the encoding time to rise. A maximum is reached when these two forces balance, after which the linear decrease in $n$ becomes the dominant factor, causing the overall computation time to decrease for high levels of set difference.

Furthermore, the encoding time is directly impacted by the false positive rate, $\epsilon$, of the initial Bloom filter. A higher $\epsilon$ results in more false positives, thus increasing both $d$ and $n$, and consequently shifting the entire encoding time curve upwards. This behavior also explains the performance of the Optimal SBF. For small differences, its optimal $\epsilon$ is high, aligning its performance with the high-$\epsilon$ SBF curves. As differences increase, its optimal $\epsilon$ decreases, and its curve converges toward the performance of low-$\epsilon$ SBFs. The Rateless Bloom Filter (RBF) exhibits the same adaptive behavior, as its effective false positive rate decreases as the filter becomes larger for higher set differences. Consequently, the RBF consistently approximates the performance of the idealized Optimal SBF.

\paragraph{Decoding Time}
The decoding time, presented in Figure \ref{fig:sbf_vs_rbf_decoding_time}, is a composite metric reflecting two primary computational costs: testing all elements against the received Bloom filter and the subsequent decoding of the Rateless IBLT. The complexity of decoding the Rateless IBLT is $O(d \log d)$, where $d$ is the number of false positives remaining after the Bloom filter stage. Unlike Bloom filter operations whose cost is tied to the total set cardinality, the IBLT decoding cost is proportional only to the number of differences, which can be significantly smaller than the full set cardinality.

For static configurations, a higher false positive rate ($\epsilon$) corresponds to a smaller Bloom filter and fewer hash functions, leading to a lower computational cost for the initial Bloom filter operations. However, as the number of differences grows, the number of false positives, $d$, also increases, and the cost of decoding the IBLT becomes the dominant factor. The lower the false positive rate, the fewer the false positives, and as a result, the decoding cost increases less for lower false positive rates. A particularly interesting observation is the behavior of the $\epsilon=1\%$ configuration, where the decoding cost remains nearly constant. This is because the increasing cost of IBLT decoding is balanced by a decreasing cost of testing elements against the second-stage Bloom filter, whose size shrinks as the number of elements that test positive in the first Bloom filter decreases. The Optimal SBF's decoding time follows the lowest decoding time of the three static configurations. This is because the optimal false positive rate for transmitted data and decoding time both follow the same pattern: it starts at high false positive rates and decreases as the number of differences increases.

The decoding time of our Rateless Bloom Filter (RBF), presented in Figure \ref{fig:sbf_vs_rbf_decoding_time}, exhibits a non-monotonic behavior, initially increasing and then decreasing. This decoding time is characterized by a notably higher absolute cost, particularly for a low number of differences. We hypothesize this is due to the computational overhead associated with the iterative nature of the rateless approach, which has less locality compared to a standard Bloom filter's single-pass partitioning. For small set differences, where the $O(d \log d)$ decoding cost of the Rateless IBLT is minimal, the difference between the Bloom filter decoding times becomes more pronounced. A similar overhead is likely present in the encoding phase but is less noticeable as that operation is dominated by the asymptotically more expensive $O(n \log d)$ complexity of constructing the IBLT.

The complex shape of the RBF decoding curve is a result of a larger decrease in the cost of the second Bloom filter operation, which outweighs the concurrent increase in the Rateless IBLT decoding cost. As the number of differences increases, the $O(d \log d)$ factor becomes the dominant component for both RBF and Optimal SBF, causing their decoding times to converge. Ultimately, since encoding time is the most expensive operation for all configurations, the RBF's larger decoding cost has a limited impact on the total computation time. 

\subsubsection{Conclusion}
In conclusion, our Rateless Bloom Filter offers an adaptive and robust solution for set reconciliation. It consistently traces the theoretical minimum of an optimally configured SBF, avoiding the suboptimal communication costs of misconfigured Bloom filters without requiring any set difference cardinality estimations or parameterization. While the RBF does incur a slightly larger computational time, this overhead has a limited impact on the total computation time. Standard Bloom filters remain a viable option in specific scenarios where the set difference cardinality can be efficiently and accurately estimated (e.g., if one set is a subset of the other), allowing the use of an optimally configured SBF. Furthermore, if the number of differences is known to be very large (e.g., Jaccard index $\le 50\%$), the communication cost of the $\epsilon=1\%$ configuration is practically the same as that of the RBF and the Optimal SBF, making it a competitive choice.

\subsection{Hybrid Rateless Set Reconciliation vs State-Of-The-Art}
We now compare our proposed Hybrid Rateless Set Reconciliation protocol (RBF + RIBLT) against three state-of-the-art algorithms: PinSketch \cite{pinsketch}, Parity Bitmap Sketch (PBS) \cite{pbs}, and Rateless IBLTs (RIBLT) \cite{rateless_set_reconciliation}.

We select these methods because they represent the state-of-the-art protocols operating on fixed-size elements:
\begin{itemize}
    \item \textbf{PinSketch} is the algorithm that represents the minimal (fixed sized) communication cost. For fixed-size elements, the communication cost corresponds to the number of differing elements times the element size, achieving an optimal communication complexity of $\mathcal{O}(d)$. However, its high decoding complexity of $\mathcal{O}(d^2)$ makes it generally untractable for practical use cases where $d$ is not guaranteed to be small.
    \item \textbf{PBS} has a higher communication overhead than PinSketch but maintains an $\mathcal{O}(d)$ communication complexity. Crucially, it improves tractability with a linear decoding complexity of $\mathcal{O}(d)$. PBS is an interactive protocol that can require multiple communication rounds, introducing additional latency.
    \item \textbf{RIBLT} (Rateless IBLT) also exhibits a higher communication overhead than PinSketch, maintaining an $\mathcal{O}(d)$ complexity. Its decoding complexity is $\mathcal{O}(d \log d)$, which, while larger than PBS, is still tractable for large $d$ (unlike $\mathcal{O}(d^2)$).
\end{itemize}
A significant practical limitation shared by both PinSketch and PBS is the mandatory requirement for an accurate, pre-estimated value of the set difference cardinality $d$ to correctly parameterize the algorithm. Our Hybrid RBF + RIBLT protocol, along with the pure RIBLT, eliminates this requirement entirely, making them adaptive to unknown set divergence.

PinSketch, PBS, and RIBLT are all $\mathcal{O}(d)$ protocols for metadata cost. In contrast, our hybrid RBF + RIBLT approach, which involves the exchange of Bloom filters, introduces an $\mathcal{O}(n)$ component, where $n$ is the total set cardinality. However, the small constants present in Bloom filters make them competitive with the state-of-the-art methods when $d$ is not extremely small compared with $n$.

As just explained, PinSketch and PBS require an estimate for $d$. Additionally, PBS requires parameter tuning based on this estimate. We define the experimental conditions as follows:
\begin{itemize}
    \item \textbf{PinSketch} is included solely as a communication baseline representing the theoretical minimum. Due to its untractable $\mathcal{O}(d^2)$ decoding complexity, this algorithm is not reasonable for the use case we focus on in this work. In order to evaluate how Hybrid Rateless Set Reconciliation performs against the minimum possible metadata size when synchronizing the fixed-size digests, in this experiment we assume an oracle provides the algorithm with the exact value of $d$.
    \item \textbf{PBS} is simulated as it would be used in a real-world setting, following the approach in the PBS experiments by leveraging it's open source implementation. We use a ToW estimator to obtain the estimate for $d$ and performed parameter tuning consistent with the original PBS experiments (ensuring a $99\%$ probability of finishing within 3 rounds, with 5 differences per group on average).
\end{itemize}

\subsubsection{Full range of similarity}
\begin{figure*}
    \centering
    \begin{subfigure}{0.24\linewidth}
        \centering
        \includegraphics[width=\linewidth]{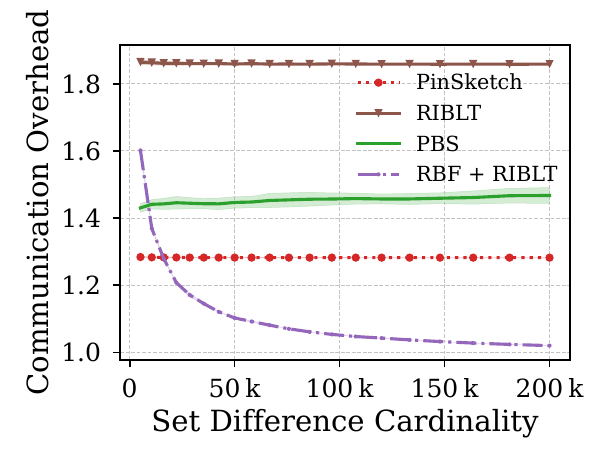}
        \caption{Communication Overhead}
        \label{fig:sota_full_range_communication_overhead}
    \end{subfigure}%
    \begin{subfigure}{0.24\linewidth}
        \centering
        \includegraphics[width=\linewidth]{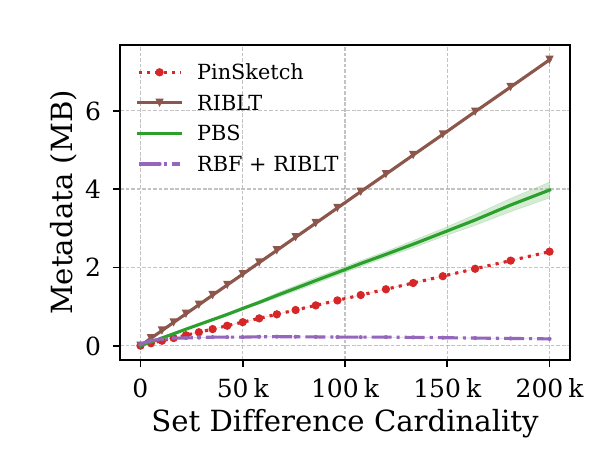}
        \caption{Transmitted metadata}
        \label{fig:sota_full_range_metadata}
    \end{subfigure}
    \begin{subfigure}{0.24\linewidth}
        \centering
        \includegraphics[width=\linewidth]{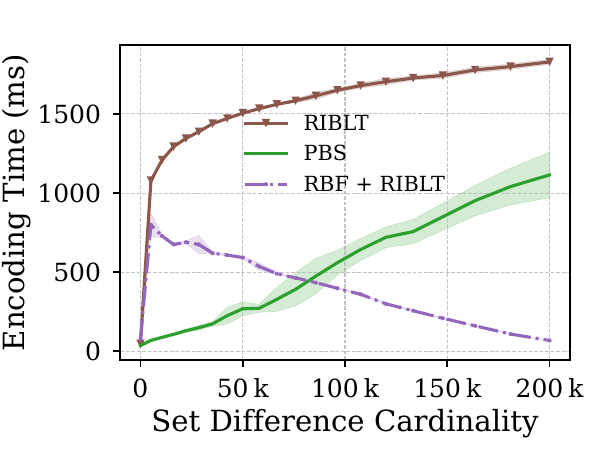}
        \caption{Encoding Time}
        \label{fig:sota_full_range_encoding_time}
    \end{subfigure}%
    \begin{subfigure}{0.24\linewidth}
        \centering
        \includegraphics[width=\linewidth]{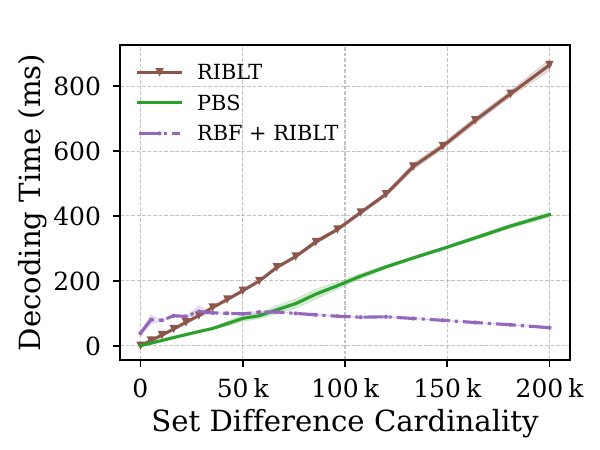}
        \caption{Decoding Time}
        \label{fig:sota_full_range_decoding_time}
    \end{subfigure}
    \caption{Hybrid Rateless Set Reconciliation vs State-Of-The-Art: Full range}
    \label{fig:sota_full_range}
\end{figure*}

\paragraph{Transmitted Metadata and Communication Overhead}
Figures \ref{fig:sota_full_range_metadata} and \ref{fig:sota_full_range_communication_overhead} present the communication performance. Focusing first on the three $\mathcal{O}(d)$ SOTA algorithms, their Communication Overhead remains practically constant, independent of $d$. This is expected as the metadata required per recovered difference does not vary significantly. In PinSketch, the metadata is the size of one digest per difference. For Rateless IBLTs, the expected metadata size is between $1.35d$ and $1.72d$ symbols (peaking at smaller $d$ values and converging to $1.35d$), and since the overhead is normalized by the minimal communication possible, it appears constant.

As expected, PinSketch has the lowest communication overhead among the SOTA methods. Between RIBLT and PBS, RIBLT has the higher overhead. The expected number of symbols to synchronize $d$ differences is $1.35-1.72d$. Furthermore, each symbol requires a fixed-size \texttt{hashSum} and a \texttt{count} field. In our implementation, the \texttt{idSum} (digest) and \texttt{hashSum} are both 64 bits to ensure low collision probability. Even ignoring the 64-bit \texttt{count} field (which could potentially be smaller), the expected communication overhead of RIBLTs is $2.7-3.48\times$ larger than PinSketch. PBS has a lower communication overhead, measured as $1.54-2.08\times$ higher than PinSketch \cite{pbs}.

Hybrid Rateless Set Reconciliation (RBF + RIBLT), due to its $\mathcal{O}(N)$ Bloom filter component, initially has a high communication overhead. In fact, for the highest similarities ($95\% \le \text{similarity} \le 100\%$), our protocol's overhead is even higher than that of RIBLT. The next experiment will focus on this region of high similarity to understand this trade-off.

However, over the total spectrum of differences, the small constants of the Bloom filter allow the RBF + RIBLT to scale very efficiently. When the similarity is lower than $85\%$ ($d \approx 18\text{k}$), Hybrid Rateless Set Reconciliation becomes the most communication-efficient protocol, outperforming even the oracle-parameterized PinSketch baseline, the minimum for fixed-size element set reconciliation algorithms. we will show that PinSketch is too computationally expensive to be used in practice even for moderate values of $d$. Thus, RBF + RIBLT results in the least communication cost among all computationally tractable algorithms for a similarity lower than $90\%$ ($d \approx 12\text{k}$). Additionally, it maintains the practical advantage over PBS (the second best) of not requiring any estimation of $d$ or parameter tuning.

Figure \ref{fig:sota_full_range_metadata} clearly illustrates the difference in metadata transmission: RBF + RIBLT leads to much smaller metadata transmissions as the number of differences grows compared to the state-of-the-art protocols, achieving up to a $92\%$ reduction in metadata transfer. Note that these four reconciliation algorithms are ultimately tasked with transmitting the same differing state (the elements themselves). Hence, the only variable cost we can minimize is the metadata required to identify these differences, a cost which our algorithm reduces by up to $92\%$.

\begin{figure}
    \centering
    \includegraphics[width=0.9\columnwidth]{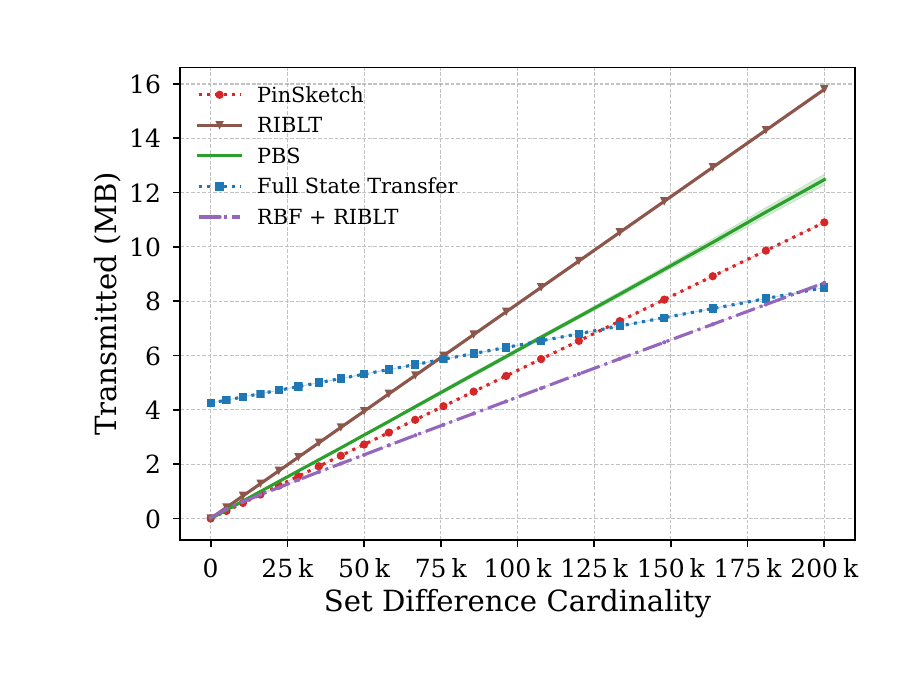}
    \caption{Hybrid Rateless Set Reconciliation vs State-Of-The-Art: Full range - Transmitted total}
    \label{fig:sota_full_range_total}
\end{figure}
Figure \ref{fig:sota_full_range_total} shows how this impacts the total transmitted data, which includes the fixed cost of the differing variable-sized elements. For comparison, we also plot the performance of a Full State Transfer algorithm, where Replica A sends its entire state to B, which then computes and sends back the missing differences. The purpose of including this baseline is to evaluate whether reconciliation protocols remain advantageous for very large differences.

The total transmitted data analysis leads to two main observations:
First, Hybrid Rateless Set Reconciliation leads to a reduction in transmitted data of approximately $20\%$ for large differences compared to PinSketch, the state-of-the-art minimum for synchronizing fixed-size element digests. This reduction is larger than $30\%$ when compared to the computationally tractable SOTA algorithms (PBS and RIBLT). As noted previously, RBF + RIBLT also performs best for similarities below $85\%$.

Second, we observe how the algorithms compare to Full State Transfer. PBS results in more total data transfer than Full State Transfer for similarities below $30\%$, while for PinSketch, the crossover point is at $20\%$ similarity. For Hybrid Rateless Set Reconciliation, Full State Transfer is only more efficient for the very lowest similarities, and even in those cases, the additional overhead of RBF + RIBLT is very low, performing practically equally.

\paragraph{Computation Time}
The computational cost of the reconciliation protocols is analyzed through Encoding Time (Figure \ref{fig:sota_full_range_encoding_time}) and Decoding Time (Figure \ref{fig:sota_full_range_decoding_time}). We do not include the computation time for PinSketch in these full-range graphs because its $\mathcal{O}(d^2)$ complexity makes it untractable for large values of $d$, as we will demonstrate in the high-similarity experiment.

The Encoding Time in Figure \ref{fig:sota_full_range_encoding_time} shows that pure RIBLT has a higher cost than PBS. However, a key advantage of RIBLT is that its encoding process can be amortized over set updates: since the RIBLT relies on a universal sequence of symbols, a large prefix can be precomputed and maintained by inserting or deleting elements as the set is updated. In contrast, the data structures in PBS must be built using parameters dependent on the specific estimate of $d$ for a given reconciliation attempt, meaning they cannot be precomputed. Furthermore, the PBS encoding time shown does not account for the computation time necessary for obtaining the estimate of $d$ or for parameter tuning, in accordance with the original PBS experiments whose source code we leveraged.

The encoding time for Hybrid Rateless Set Reconciliation (RBF + RIBLT) is initially higher than PBS, but as the number of differences becomes large, RBF + RIBLT becomes more efficient than PBS. For large differences, RBF + RIBLT offers both less encoding time and a lower communication cost than PBS. While comparing pure RIBLT and RBF + RIBLT encoding time is less relevant if RIBLT is amortized, we can see that if encoding cannot be amortized, pure RIBLT exhibits a high encoding time, consistently larger than RBF + RIBLT.

For Decoding Times (Figure \ref{fig:sota_full_range_decoding_time}), PBS's linear complexity $\mathcal{O}(d)$ is theoretically faster than RIBLT's $\mathcal{O}(d \log d)$, which is reflected in the graph. The decoding time for RBF + RIBLT is initially higher than PBS but becomes smaller as $d$ grows large. This once again confirms that for large $d$, RBF + RIBLT is superior, being both computationally less expensive and having a lower communication cost than PBS.

If we consider the total computation time and amortize RIBLT's encoding cost (treating it as 0), the total computation time for RIBLTs would be smaller than PBS's total cost. The total computation time for RBF + RIBLT is initially higher than both PBS and amortized RIBLT but becomes the lowest as $d$ increases.

\subsubsection{High similarity}
In this experiment, we focus on similarities between 85\% and 100\% with 0.5\% increments. This is precisely the range of similarities in which hybrid rateless set reconciliation is not the best performing algorithm, and the purpose of this experiment is to visualize what the overhead is.

\begin{figure*}
    \centering
    \begin{subfigure}{0.24\linewidth}
        \centering
        \includegraphics[width=\linewidth]{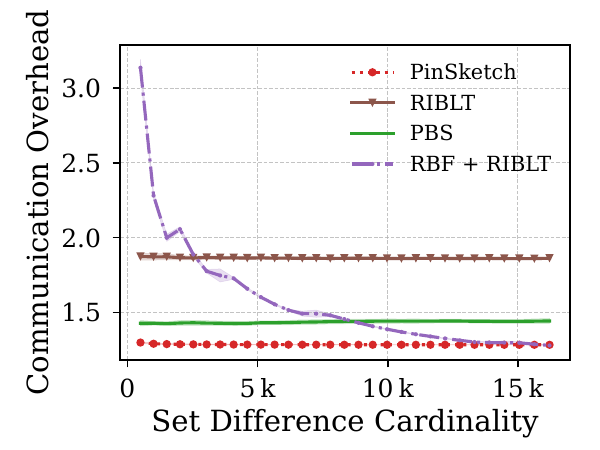}
        \caption{Communication Overhead}
        \label{fig:sota_high_similarity_communication_overhead}
    \end{subfigure}%
    \begin{subfigure}{0.24\linewidth}
        \centering
        \includegraphics[width=\linewidth]{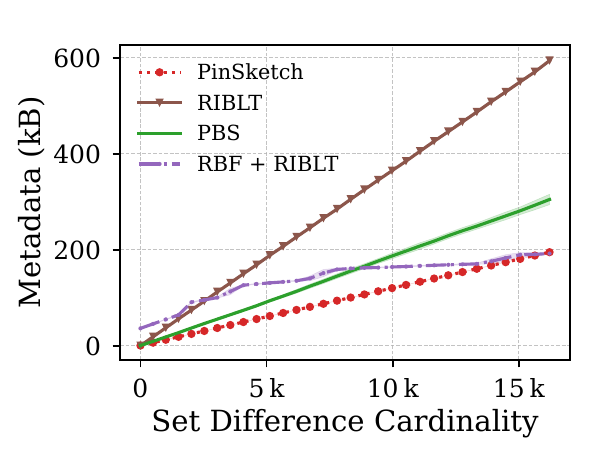}
        \caption{Transmitted metadata}
        \label{fig:sota_high_similarity_metadata}
    \end{subfigure}
    \begin{subfigure}{0.24\linewidth}
        \centering
        \includegraphics[width=\linewidth]{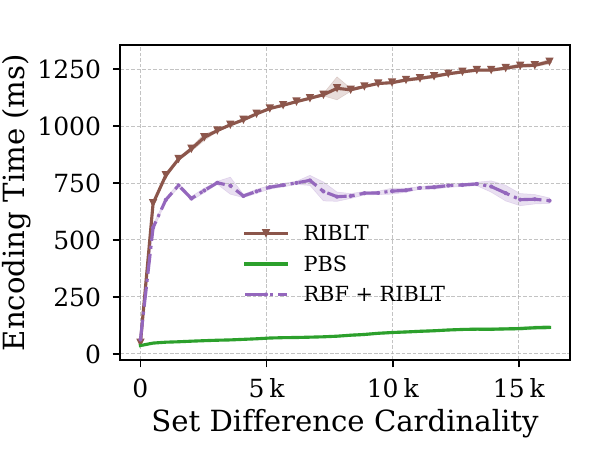}
        \caption{Encoding Time}
        \label{fig:sota_high_similarity_encoding_time}
    \end{subfigure}%
    \begin{subfigure}{0.24\linewidth}
        \centering
        \includegraphics[width=\linewidth]{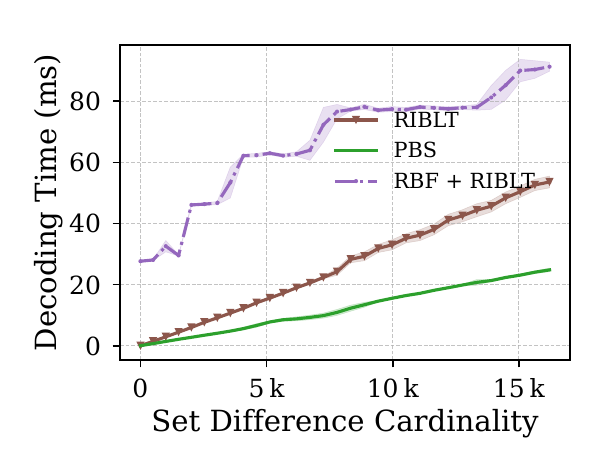}
        \caption{Decoding Time}
        \label{fig:sota_high_similarity_decoding_time}
    \end{subfigure}
    \caption{Hybrid Rateless Set Reconciliation vs State-Of-The-Art: High Similarity}
    \label{fig:sota_high_similarity}
\end{figure*}

\paragraph{Transmitted Metadata and Communication Overhead}

Figure \ref{fig:sota_high_similarity_communication_overhead} shows communication overhead. The comparison between PinSketch, PBS, and RIBLT in terms of data transfer is largely independent of the experiment range. The focus here is on the trade-off introduced by RBF + RIBLT. This figure confirms that for very small $d$ (e.g. corresponding to a similarity of $0.5\%$), the communication overhead for RBF + RIBLT is significantly higher than state-of-the-art methods. This initial overhead is due to transmitting the Bloom filter of size $\mathcal{O}(N)$, where $N$ is much larger than the small $d$.

Figure \ref{fig:sota_high_similarity_metadata} shows the metadata costs. When synchronizing identical states ($d=0$), RBF + RIBLT has a much larger (percentage-wise) metadata cost, sending close to $50\text{KB}$ while other algorithms require only a few bytes to detect convergence. This makes it clear that RBF + RIBLT is not the ideal choice for anti-entropy protocols where extremely small differences (if any) are expected between frequent synchronization rounds.

However, as $d$ increases, the introduction of RBFs quickly becomes beneficial. For similarities below $97.5\%$ ($d \approx 2.5\text{k}$), RBF + RIBLT is already more communication-efficient than pure RIBLT. Furthermore, as was shown in the full-range experiment, RBF + RIBLT outperforms PBS below $90\%$ similarity and then PinSketch below $85\%$ similarity. For the critical use case of synchronizing potentially very dissimilar states (e.g., bringing an outdated replica back online after a network partition), the additional cost of using RBF + RIBLT is never more than approximately $75\text{KB}$ compared to the best-performing SOTA algorithm in this regime (PinSketch). This is dramatically better than the additional $2\text{MB}$ of metadata transfer incurred by PinSketch when compared to RBF + RIBLT at completely dissimilar states.

This adaptability is the core advantage of RBF + RIBLT. If a general estimate of $d$ (e.g., extremely small, small, moderate, or high) is known, one can choose the best performing algorithm for that range. If, however, one must choose a single algorithm to synchronize states of unknown and possibly high divergence, RBF + RIBLT provides excellent adaptability, achieving the best overall worst-case communication overhead.

\paragraph{Decoding Time with PinSketch}
\begin{figure}
    \centering
    \includegraphics[width=0.48\columnwidth]{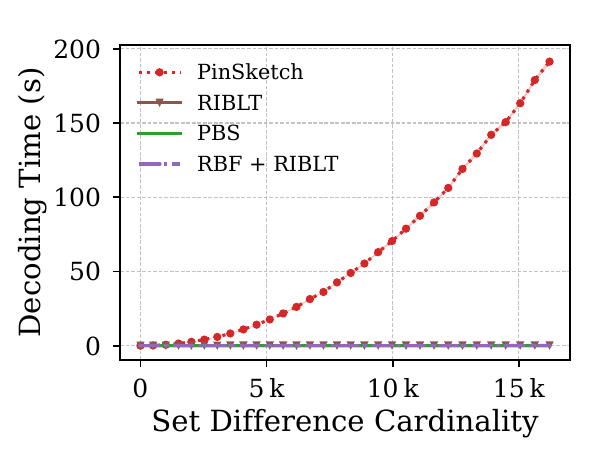}
    \caption{Hybrid Rateless Set Reconciliation vs State-Of-The-Art - Decoding Time with PinSketch}
    \label{fig:sota_high_similarity_decoding_time_pinsketch}
\end{figure}

Figure \ref{fig:sota_high_similarity_decoding_time_pinsketch} shows the decoding times including  PinSketch. The graph demonstrates the intractability of PinSketch's $\mathcal{O}(d^2)$ decoding complexity, making the decoding time of all other algorithms appear negligible. For a similarity of $90\%$ ($d \approx 12.5\text{k}$), PinSketch's decoding time is already approximately $100\text{s}$. In most practical use cases, the network could transfer orders of magnitude more data in $100\text{s}$ than the total transmitted data difference between PinSketch and the worst-performing communication protocol. We conclude that PinSketch is only practical if the number of differences is extremely small, such as in highly frequent anti-entropy protocols with a low update rate, which is not the primary use case we address in this work.

\paragraph{Computation Time without PinSketch}
We now analyze the encoding and decoding times for the three computationally tractable algorithms. Figure \ref{fig:sota_high_similarity_encoding_time} shows that PBS has a significantly lower encoding time, ranging between $0$ and $100\text{ms}$. RBF + RIBLT has an encoding time mostly between $600\text{ms}$ and $800\text{ms}$ after an initial jump. This jump, also seen in pure RIBLT, is due to the first symbols having the highest encoding time per symbol (as the probability of mapping to symbol $i$ is $\mathcal{O}(1/i)$), causing a sharp increase when $d$ moves from zero to a positive value. As before, the encoding time for pure RIBLT is higher than both RBF + RIBLT and PBS, but its cost can be amortized on updates.

The Decoding Time (Figure \ref{fig:sota_high_similarity_decoding_time}) is the lowest for PBS ($\mathcal{O}(d)$), followed by RIBLT ($\mathcal{O}(d \log d)$), and then RBF + RIBLT. We also note that the computation time is largely dominated by the encoding time, which is almost an order of magnitude greater than the corresponding decoding time for each algorithm and similarity level.

Considering the trade-offs, the value of $d$ where Hybrid Rateless Set Reconciliation becomes the preferred choice over PBS depends on the specific CPU and network capabilities of the deployment environment. For similarities above $90\%$, PBS has both lower computational time and lower communication cost. For similarities between $60\%$ and $90\%$, PBS maintains a lower computational time but has a higher communication cost. For similarities below $60\%$, RBF + RIBLT is superior in terms of both communication cost and computation cost. As the similarity decreases from $90\%$ to $60\%$, the computational difference decreases while the communication difference in favor of RBF + RIBLT increases, but the exact crossover point remains hardware dependent.

\section{Related Work}
\label{sec:related_work}

Set reconciliation techniques are fundamentally characterized by an inherent trade-off between communication overhead and computational complexity. Early methods like PinSketch \cite{pinsketch}, which leverage error-correcting codes by treating the differing elements as errors, achieve near-optimal $O(d)$ communication (where $d$ is the difference cardinality) but suffer from high $O(d^2)$ decoding complexity. The Invertible Bloom Lookup Table (IBLT) \cite{iblt} solved this computational bottleneck with efficient $O(d)$ encoding and decoding. However, IBLTs are probabilistic and must be significantly overprovisioned to ensure a low failure rate, increasing the overall communication cost. The Parity Bitmap Sketch (PBS) \cite{pbs} balances PinSketch's communication efficiency with IBLT's low O(d) computational complexity. It achieves this by employing a divide-and-conquer strategy that partitions sets with large differences into numerous subsets (each with an expected $O(1)$ difference cardinality), making the otherwise high $O(d^2)$ decoding complexity of the underlying error-correcting codes computationally tractable. Graphene \cite{graphene} uses a combination of Bloom Filters (BFs) and IBLTs, where the IBLT corrects BF false positives. The combined communication cost is lower than using either of the data structures in isolation.

A critical, shared limitation of PinSketch, IBLTs, and PBS is the necessity of estimating the difference cardinality $\mathbf{d}$ prior to reconciliation. This requirement results in an initial communication overhead for estimation and risks protocol failure if the sketch is insufficiently overprovisioned. Graphene \cite{graphene} estimates $d$ without requiring an additional communication step by assuming the transmitter's set is a subset of the receiver's. This assumption is valid in their specific domain, but it makes Graphene unsuitable for general symmetric difference reconciliation.

The challenge of parameter estimation is fundamentally addressed by rateless techniques. Rateless IBLTs (RIBLTs) \cite{rateless_set_reconciliation} extend the concept by streaming a coded sequence until successful decoding is achieved, thereby eliminating the need for a $\mathbf{d}$ estimate. ConflictSync \cite{conflictsync} uses a BF/RIBLT combination to overcome the high communication cost of RIBLTs for large $d$, but it lacks a robust methodology for correctly parameterizing the initial Bloom Filters.

The idea of streaming streaming small BFs \cite{low_complexity_set_reconciliation} was previously introduced in the context of set reconciliation. However, BFs are exchanged interactively until convergence is achieved, lacking a mechanism to efficiently transition to a robust $\mathbf{O(d)}$ set reconciliation method when the BFs exhibit diminishing returns.

\section{Conclusion and Future Work}
\label{sec:conclusion}

Previous research on set reconciliation has often neglected the challenge of variable-sized elements, stating that the problem can be reduced to synchronizing fixed-size element digests. In contrast, this work demonstrates the significant practical benefits of designing reconciliation protocols tailored for variable-sized elements. Our results show that by leveraging Bloom Filters within our hybrid approach, we achieve a communication cost reduction of over $20\%$ compared to PinSketch, the state-of-the-art method for lowest communication complexity. When considering only algorithms that are computationally tractable with large differences, this reduction increases to over $30\%$. We also introduce Rateless Bloom Filters (RBFs), a novel adaptive Bloom filter that closely matches the performance of an optimally configured standard Bloom Filter, yet requires no prior parametrization or estimate of the unknown set difference cardinality $\mathbf{d}$.

Despite these advances, our current approach is suboptimal for replicas with extremely high similarity (Jaccard index above $85\%$), as initial RBF slices consume excessive bandwidth relative to the small difference $\mathbf{d}$. An interesting area of future work is inverting the protocol: initiating reconciliation with RIBLTs. From observing a given prefix, we can analyze the cell counts to derive a statistical estimate of the set intersection. If the divergence is expected to be large, this allows us to efficiently switch to RBFs.

\clearpage

\bibliographystyle{ACM-Reference-Format}
\bibliography{sample}

\end{document}